\documentclass[twocolumn,twoside,slac_two]{revtex4}
\usepackage{graphicx}
\usepackage{fancyhdr}
\usepackage{graphics}
\usepackage{epstopdf}
\usepackage{textpos}
\newcommand{\ket}[1]{|#1\rangle}

\begin{document}

\title{\centering Progress on the neutrino mixing angle, $\theta_{13}$}
\author{
\centering
\includegraphics{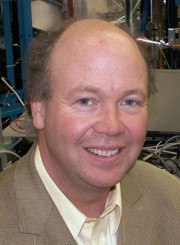} \\
\begin{center}
D. Karlen
\end{center}}
\affiliation{\centering University of Victoria and TRIUMF, British Columbia, Canada}
\begin{abstract}
Until recently, measurements of $\theta_{13}$, which describes the $\nu_e$ component in the $\nu_3$ mass eigenstate, gave only
upper limits, leaving open the possibility that it is zero and eliminating a source of CP violation in the neutrino sector.
This year has seen significant advances in measuring $\theta_{13}$ as
precision reactor experiments move from construction phase to physics operation and
accelerator experiments give first indications that $\theta_{13}$ differs from zero.
In the years to come, the results from these and other experiments will stringently test the PMNS framework for neutrino oscillation
and will start to give information about neutrino CP violation and the neutrino mass hierarchy.
This paper summarizes the situation for measuring $\theta_{13}$ at this pivotal time in neutrino research.
\end{abstract}

\maketitle
\thispagestyle{fancy}


\section{INTRODUCTION}

It has been a little more than a decade since experiments firmly established oscillations in the neutrino sector by measuring interaction rates of atmospheric neutrinos produced by cosmic rays and solar neutrinos.\cite{ref:SKosc,ref:SNOosc}
This behaviour is summarized in the PMNS (Pontecorvo, Maki, Nakagawa, and Sakata) framework with the states of
definite flavour being a linear combination of states of definite mass, and parameterized by the unitary
 PMNS matrix, $U$.\cite{ref:MNS,ref:P}
\begin{eqnarray*}
\ket{\nu_\alpha}&=&\sum_{i=1}^{3}U^*_{\alpha i}\ket{\nu_i} \ \ \ \ \alpha=e,\mu,\tau \\
U&=&\left[\begin{array}{ccc}
1 &  0 &  0      \\
0 & c_{23} & s_{23}     \\
0  & -s_{23}  &   c_{23} \\
	\end{array}\right] \times
\left[\begin{array}{ccc}
c_{13} &  0 &  s_{13}e^{-i\delta}      \\
0 & 1 & 0     \\
-s_{13}e^{-i\delta}   & 0  &   c_{13} \\
	\end{array}\right]   \\
&\times&
\left[\begin{array}{ccc}
c_{12} &  s_{12} &  0      \\
-s_{12} & c_{12} & 0    \\
0  & 0  &   1 \\
	\end{array}\right] \ \ \  \ \ \ 
\begin{array}{ll}
s_{13}\equiv\sin\theta_{13} \\
c_{13}\equiv\cos\theta_{13}
\end{array}
\end{eqnarray*}

As a neutrino propagates, the phases of its components will change relative to each other, provided the mass eigenstates are non-degenerate.
This leads to the possibility that a neutrino interaction will produce a charged
lepton of different flavour than the one associated with its production, a process known as neutrino oscillation.
The probability for a neutrino to oscillate into a different flavour depends on the PMNS matrix elements, the differences between
the squared masses of the eigenstates, the neutrino energy, and the distance it has traveled.

The PMNS angles describing solar ($\theta_{12}$) and atmospheric ($\theta_{23}$) neutrino oscillation have been well measured, as have
been the differences between the masses of the eigenstates.\cite{ref:PDG}
\begin{eqnarray*}
\sin^2 2\theta_{12}&=&0.861^{+0.026}_{-0.022} \\ 
\Delta m^2_{21} &=& (7.59\pm0.21) \times 10^{-5}\,\textrm{eV}^2 \\ 
\sin^2 2\theta_{23} > 0.92 \ \
\vert\Delta m^2_{32}\vert &=& (2.43\pm0.13) \times 10^{-3}\,\textrm{eV}^2 \ \ 
\end{eqnarray*}

Until this year, the best single constraint on $\theta_{13}$ 
came from the CHOOZ experiment, $\sin^2 2\theta_{13} < 0.15$ at 90\% C.L.\cite{ref:CHOOZ}
If the angle is zero, perhaps due to some symmetry, this would eliminate a potential source of CP violation
in the lepton sector, parameterized by $\delta$ in the equation above.
As CP violation is an essential ingredient in the explanation of the matter-antimatter asymmetry in the Universe, the
question of whether this angle is non-zero is of great interest.
Should it differ from zero, its magnitude determines the feasibility of future experiments to measure $\delta$.

In the following sections, recent progress towards improving our understanding of $\theta_{13}$ by reactor and accelerator experiments is reviewed.

\section{REACTOR EXPERIMENTS}

Neutrinos were discovered in an experiment at the Savannah River nuclear reactor in 1956 through
the inverse beta decay (IBD) reaction, $\overline{\nu}_e + p \rightarrow e^+ + n$.\cite{ref:Reines}
The positron annihilation produces a prompt signal which is followed later by a signal from the neutron capture.
The experiment used two 200$\,$L water tanks with cadmium salt added to reduce the neutron capture delay to less than 10$\,\mu$s.
The two tanks were sandwiched by three larger scintillator tanks for detecting the gamma rays, and a delayed coincidence trigger was used.

In the 1980s and 1990s, several similar experiments were set up to measure the spectrum of neutrinos involved in
IBD reactions at various distances 
from reactors at ILL, Bugey, Rovno, Goesgen, Krasnoyark, Palo Verde, and Chooz.
In absense of oscillations, the product of the rapidly falling neutrino flux and the rapidly growing IBD cross section with energy,
would result in an observed spectrum of neutrinos that is peaked between 3 and 4 MeV.
At a suitable baseline distance, experiments can reveal neutrino oscillation
by observing a distortion in the spectrum of neutrinos interacting via the IBD reaction.

The disappearance probability for reactor neutrinos travelling a distance $L$ is approximated by
\begin{eqnarray*}
1-P(\overline{\nu}_e\rightarrow\overline{\nu}_e) \approx
\sin^2 2\theta_{13} \sin^2  \left({\Delta m^2_{31} L}\over{4E_\nu}\right) \\
+\cos^4\theta_{13}\sin^2 2\theta_{12} \sin^2 \left({\Delta m^2_{21} L}\over{4E_\nu}\right).
\end{eqnarray*}
The first term is the more sensitive one for small values for $\theta_{13}$ and the sensitivity is maximized when
\begin{eqnarray*}
\sin^2 \left({\Delta m^2_{31} L}\over{4E_\nu}\right) \approx 1 \\
{\Delta m^2_{31} L\over{4E_\nu}} \approx {\pi\over2}
\end{eqnarray*}
\begin{equation}
L \approx 0.5\,\hbox{km}{E_\nu\over{\hbox{MeV}}}
\label{eqn:baseline}
\end{equation}
and so the optimal baseline distance is 1-2$\,$km, for observable reactor neutrinos.

To date, the most sensitive reactor measurement of $\theta_{13}$
comes from the Chooz experiment, located in Northern France.
It consisted of a 5$\,$ton target of liquid scintillator with gadolinium to enhance neutron capture with
a baseline distance of about 1$\,$km to the Chooz reactor.
From 3600 events collected in the period 1997-1998, the experiment was able to limit $\theta_{13}$
\begin{eqnarray*}
\sin^2 2\theta_{13}<0.15\  \   @  90\% CL \ \ \ \  (\theta_{13}<11^\circ)
\end{eqnarray*}
for $\Delta m^2_{13} = 2.5 \times 10^{-3}\,$eV$^2$.\cite{ref:CHOOZ}

Three new reactor experiments designed to have significantly better sensitivity for $\theta_{13}$ have begun collecting data in 2011;
Double Chooz\cite{ref:doublechooz} in France, RENO\cite{ref:reno} in Korea, and Daya Bay\cite{ref:daya} in China.
All three use larger detectors and are better shielded than the earlier experiments.
All three also include identical near detectors to reduce systematic uncertainties in modeling the neutrino
flux and cross section and the detector acceptance.
The near detectors are located at baselines of about 300-500$\,$m, where the disappearance
probability for observable reactor neutrinos is very small.

\begin{figure}
\includegraphics[width=75mm]{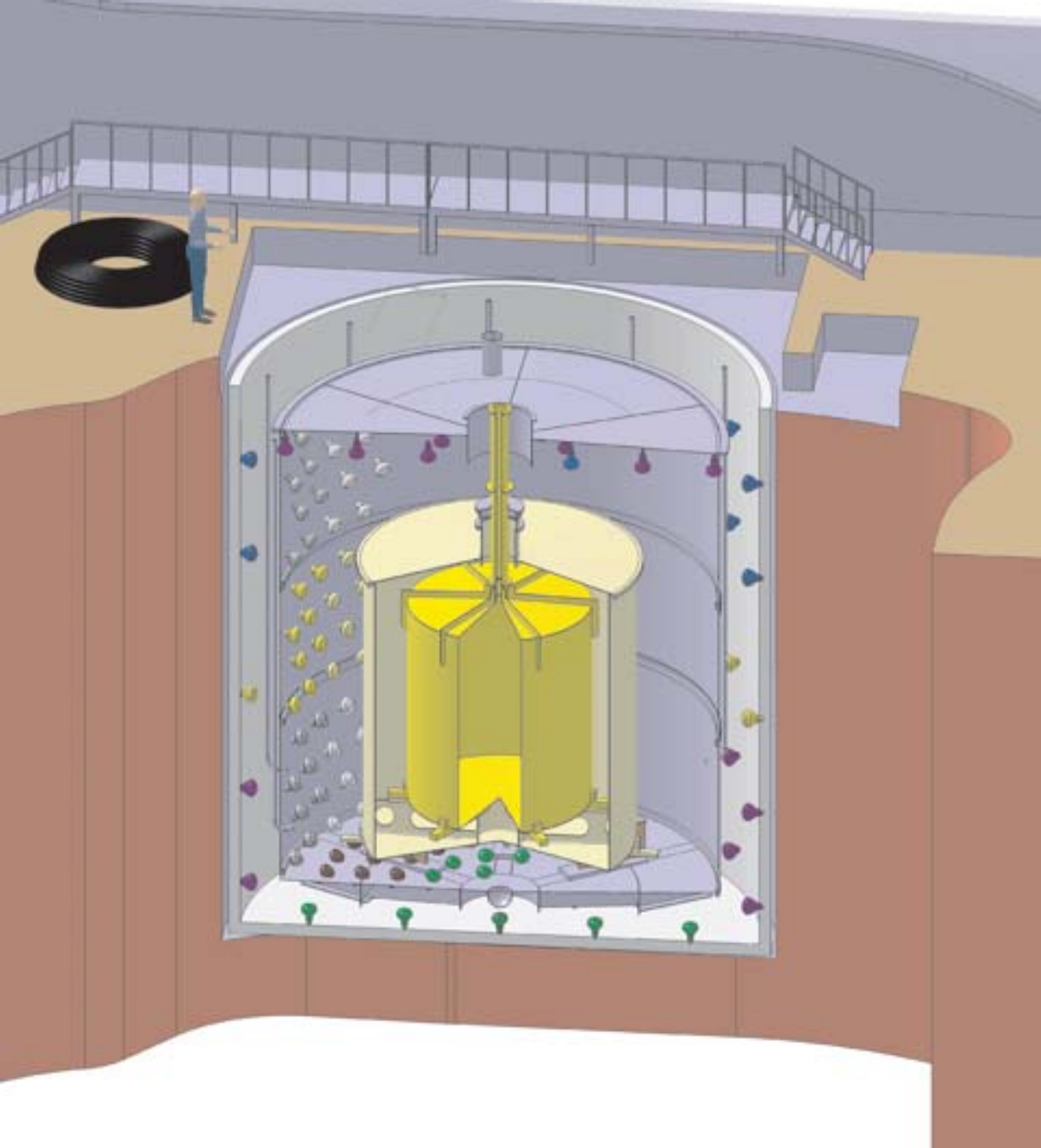}
\caption{Illustration of the Double Chooz detector. It consists of four concentric volumes - the target,
gamma-catcher, buffer, and cosmic veto.
} \label{fig:doublechooz}
\end{figure}

The Double Chooz design uses four concentric cylindrical volumes, as illustrated in Fig.~\ref{fig:doublechooz}.
Innermost is the 8.3$\,$ton target, consisting of liquid scintillator with gadolinium held in an acrylic cylinder.
The target is located within the so-called gamma-catcher, a larger acrylic cylinder filled with liquid scintillator without
gadolinium.
These are located within a buffer vessel containing non-scintillating mineral oil and 390 phototubes to measure light from the
inner scintillator volumes.
This is surrounded by liquid scintillator and phototubes that act as a cosmic muon veto.
Phase 1 operation of Double Chooz began in April 2011, with all systems for the far detector working as expected.
Phase 2 will begin in the end of 2012, when the near detector site and detectors are expected to be completed.

The two RENO detectors closely resemble the Double Chooz detectors, using four concentric cylindrical volumes.
The RENO detectors were completed in February 2011, and filled with liquids in July 2011.
Regular data-taking with both detectors began in August 2011.

\begin{table*}[t]
\begin{center}
\caption{Comparisons of the new reactor experiments. Shown are the reactor power, the baseline distance to near and far detectors,
the overburden depth in meters of water equivalent, the target mass, sensitivity to $\sin^2 2\theta_{13}$ expressed as the expected
90\% C.L.~limit if $\theta_{13}=0$ with a three year dataset, and the starting dates for data collection.}
\begin{tabular}{|l|c|c|c|c|c|c|c|}
\hline & \textbf{Power} & \textbf{L near/far} &
\textbf{Depth n/f} & \textbf{Target mass} & \textbf{$\sin^2 2\theta_{13}$} 
& \textbf{start date} & \textbf{start date}
\\
\textbf{Experiment} & \textbf{(GW)} & \textbf{(m)} & \textbf{(mwe)} & \textbf{(tons)} & \textbf{sensitivity}
& \textbf{near detector(s)} & \textbf{far detector}
\\
\hline Double Chooz & 8.7 & 400/1050 & 110/300 & 8.3/8.3 & 0.03 & end/2012 & 4/2011 \\
\hline RENO & 16.4 & 290/1380 & 120/450 & 16/16 & 0.02 & 8/2011 & 8/2011 \\
\hline Daya Bay & 17.7 & 360-1980 & 270/910 & 40,40/80 & 0.01 & 8/2011 + fall/2011 & summer/2012\\
\hline
\end{tabular}
\label{tab:reactor}
\end{center}
\end{table*}

The Daya Bay experiment consists of multiple identical detectors, with four detectors at the far site, and two detectors at each
of the two near detector locations.
The standard four concentric volume design is used, but with the detectors at the same site sharing the outer veto volume.
The large number of detectors allows the properties of the detectors to be compared to confirm systematic errors.
Tunnels connect the detector locations, which would allow the detectors to be exchanged if 
the detector systematic errors are significant.
The two detectors at one near detector site have become operational in August 2011, with the other near detector site
expected to come on-line later in 2011.
The far detector is expected to start operations in the summer of 2012.

There will be significant competition between the three new reactor experiments over the coming years.
A summary of their specifications and their starting dates for 
data collection are given in Table~\ref{tab:reactor}.
The neutrino detectors themselves are remarkably similar in design concept, which may be optimal in terms of performance for
each individual experiment.
When combining the results of these experiments in the future, it will be important to consider common systematic errors that
come from having such similar designs.

\section{ACCELERATOR EXPERIMENTS}

In 1962, the existence of a second kind of neutrino, $\nu_\mu$, was established by an experiment that directed high energy
protons from the BNL AGS onto a target.\cite{ref:AGS}
Neutrinos produced in the decay of secondary hadrons passed through 13.5$\,$m of shielding and through 
10 one-ton modules consisting of aluminum plates with spark chambers.
Particles produced within the modules, arising from neutrino interactions, were found to be penetrating
like muons, rather than showering like electrons. 

Contemporary accelerator neutrino experiments estimate $\theta_{13}$ by measuring the $\nu_e$ appearance probability
in a nearly pure $\nu_\mu$ beam,
\begin{eqnarray*}
P(\nu_\mu\rightarrow\nu_e)\approx\sin^2 \theta_{23} \sin^2 2\theta_{13} \sin^2\left({\Delta m^2_{32} L\over 4 E_\nu}\right).
\end{eqnarray*}
Since the energies of $\nu_\mu$ beams produced by accelerators are of order 1$\,$GeV, 
Eq.~\ref{eqn:baseline} indicates the optimal baseline is several hundred km.
At such great distances, very high power beams and large detectors are necessary to be sensitive to
a relatively small $\theta_{13}$.
In addition, it is necessary that the detectors can distinguish $\nu_e$ charged current interactions from other types of neutrino
interactions and that the $\nu_e$ component in the initial beam be small and well estimated.

The $\nu_e$ appearance probability contains additional terms that depend on the CP violation parameter, $\delta$, and the
sign of the mass difference $\Delta m^2_{32}$.
The reactor $\nu_e$ disappearance probability does not depend on these quantities, so the information from reactors and
accelerators are complementary.

\subsection{The T2K experiment}

The T2K project\cite{ref:t2kexperiment},
approved in December 2003, arose from a fortunate coincidence of having the two major facilities needed to measure $\theta_{13}$
separated by an appropriate distance:
\begin{itemize}
\item The Japan Proton Accelerator Research Center (JPARC) was under construction at the time, and designed to produce the world's highest
intensity proton beam, thus suitable as a driver for a long baseline neutrino experiment.
\item The Super Kamiokande detector (SK), operating for many years as the world's largest underground water Cerenkov detector,
was proven to be an excellent detector of neutrinos from solar, atmospheric, and the earlier K2K experiment, and was located at distance of
295$\,$km from JPARC.
\end{itemize}

\subsubsection{The T2K beam and detectors}

The centre of the T2K neutrino beam is intentionally directed 2.5$^\circ$ away from SK in order to increase the flux of 0.6$\,$GeV neutrinos
passing through SK (giving the largest $\nu_e$ appearance probability for a 295$\,$km baseline) and to decrease the flux of
higher energy neutrinos passing through SK (which can be mis-reconstructed as lower energy neutrinos, an important background in
the search for $\nu_e$ appearance).
Charged current interactions, which identify the neutrino type by the flavour of the produced charged lepton, are predominantly 
quasi-elastic for neutrinos below 1$\,$GeV.
This allows for a simple selection of neutrino interactions and to estimate the neutrino energy using only the energy and
direction of the charged lepton.

The T2K neutrino beam is produced by bringing 30$\,$GeV protons from the JPARC main ring onto a 1$\,$m long graphite target.
The positively charged hadrons (primarily pions) are focused by a set of 3 pulsed magnetic horns and pass through a 100$\,$m decay volume
filled with helium gas.
The majority of pions decay within that volume, producing a high purity $\nu_\mu$ beam with a small $\nu_e$ contamination from kaon and muon decays, 
and a small anti-neutrino contamination from negative hadrons.

Several detector systems are in place to monitor the beam properties and its stability.
Proton beamline monitors measure the beam current and extrapolate its position and direction on the target.
A muon monitor, located immediately downstream of the decay volume, measures the distributions of muons produced from hadron decays.
Near detectors, located 280$\,$m from the target, measure the neutrino profile and interaction rate around the beam centre 
(with the on-axis INGRID detector)
and measure the neutrino spectrum, purity, and general neutrino interaction properties along the off-axis direction to SK
(with the off-axis ND280 detector).

The on-axis INGRID detector consists of 14 identical modules, each consisting of 7 tons of iron-scintillator sandwich design.
The modules are arranged in a cross centred on and transverse to
the neutrino beam with 7 placed horizontally side-by-side and 7 stacked in the vertical direction.

The off-axis ND280 detector is a multipurpose spectrometer using the magnet built for the CERN UA1 experiment.
The rectangular volume inside the coils, 7$\,$m along along the beam direction and 3.5$\,$m$\times$3.5$\,$m transverse to the beam,
is instrumented with a suite of detectors to precisely measure and characterize neutrino interactions in the unoscillated beam.
The pi-zero detector, consisting of scintillator bars interleaved with water target layers and brass and lead sheets to enhance photon conversion,
is designed to efficiently reconstruct neutral current neutrino interactions that produce a $\pi^0$, an important background for
$\nu_e$ appearance at SK.
Immediately downstream is the tracker, consisting of two fine grained scintillator (and water) target modules sandwiches by three time
projection chambers (TPCs) to measure the charged particles from neutrino interactions.
The TPC measurements of curvature and ionization energy loss are used to determine the momenta and 
identify the types of charged particles passing through them.
Surrounding the pi-zero and tracker are electromagnetic calorimeters to measure photons and help identify electrons produced in
neutrino interactions.
Several layers of scintillator planes are instrumented in the yoke of the magnet to provide the side muon range detectors, useful for identifying muons
produced by neutrinos and cosmic rays.

The Super Kamiokande detector has been in operation since 1996.
The inner detector fiducial volume is 22.5$\,$kton of pure water and is surrounded by 11,129 50-cm diameter photomultiplier tubes to detect the
Cerenkov light produced by relativistic charged particles.
The signature of a charged lepton that starts and stops within the fiducial volume (a so-called fully contained fiducial volume event,
consistent with coming from a neutrino interaction) is a characteristic ring pattern of light.
Because electrons produce a shower of particles as they lose energy in the water, the boundary of a Cerenkov ring from an electron
is much less sharp than that of a muon.
The capability of the detector to distinguish electron and muon neutrino interactions is well understood from high statistics atmospheric
neutrino data samples.

\subsubsection{T2K data}

The appearance probability $P(\nu_\mu\rightarrow\nu_e)$ at SK is evaluated by counting the number of
$\nu_e$ interactions coincident in time with the neutrino pulse passing through SK, corrected for backgrounds and efficiency,
and dividing by the number of $\nu_\mu$ interactions expected if there were no neutrino oscillations.
To evaluate the backgrounds and neutrino beam properties, a complete simulation of the beamline and detectors is used, along with
internal data (such as the rate of $\nu_\mu$ interactions in the near detector) and external data (such as the hadron production
rate measurements from the NA61 experiment\cite{ref:NA61} and others).

T2K physics data collection commenced in March 2010.
Following the summer 2010 shutdown, proton intensities steadily increased to 145$\,$kW.
On March 11, 2011, a devastating earthquake struck the east coast of Japan, and damage to JPARC and surrounding areas
forced an extended shutdown of the project, after collecting $1.43\times10^{20}$ protons on target, 
only a few percent of the projected amount for the experiment.
Remarkable progress has been made to restore the facilities at JPARC, and the experiment plans to begin operations again in
January 2012.

The data collected by INGRID show that the neutrino beam properties were very stable,
both in terms of the $\nu_\mu$ interaction rate per proton on target, and the direction and profile of the beam.
The beam direction variation was found to be much less than the requirement of $\pm1\,$mrad.

\begin{figure}
\includegraphics[width=85mm]{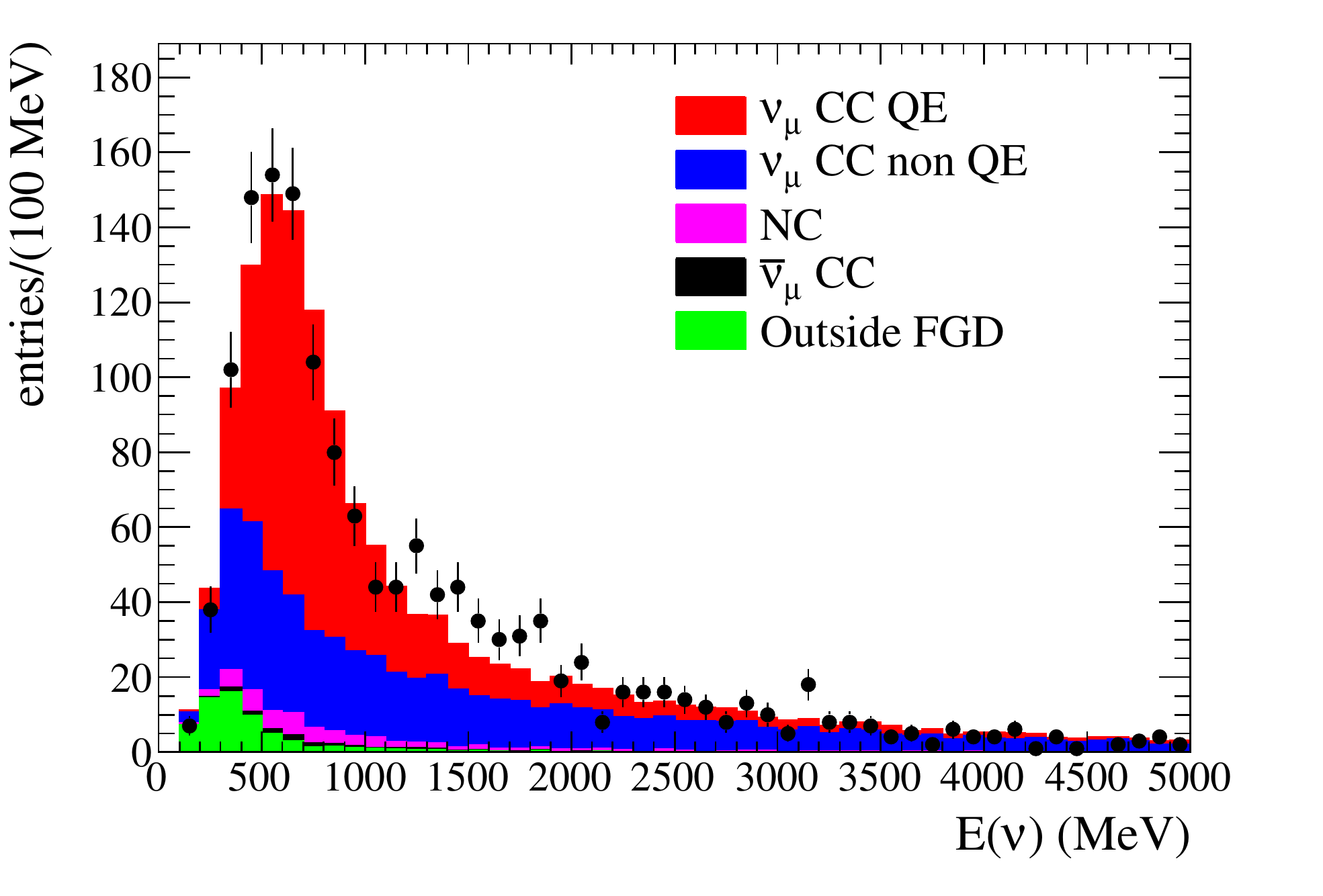}
\caption{Distribution of reconstructed $\nu_\mu$ energies for a selection of events consistent 
with $\nu_\mu$ charged current interactions in the fine grained detectors (FGD) of the off-axis ND280 tracker. The histograms indicate the expected distribution from full simulation of the experiment, with no
free parameters adjusted to improve the fit to data.
} \label{fig:nd280}
\end{figure}

The off-axis ND280 tracker measurement of charged current $\nu_\mu$ interactions were found to be in very good agreement with the T2K simulation both in terms of spectrum and rate.
The spectrum is shown in Fig.~\ref{fig:nd280} and compared to a simulation which was not tuned to the ND280 data.
The ratio of observed to simulated interaction rate is
\begin{eqnarray*}
{R^{\textrm{Data}}_{ND} / R^{\textrm{MC}}_{ND}} = 1.036\pm0.028 (\textrm{stat})^{+0.044}_{-0.037} (\textrm{det. sys})\\
\pm0.038 (\textrm{phys. model})
\end{eqnarray*}

\subsubsection{T2K result on $\nu_e$ appearance}

\begin{figure}
\includegraphics[width=80mm]{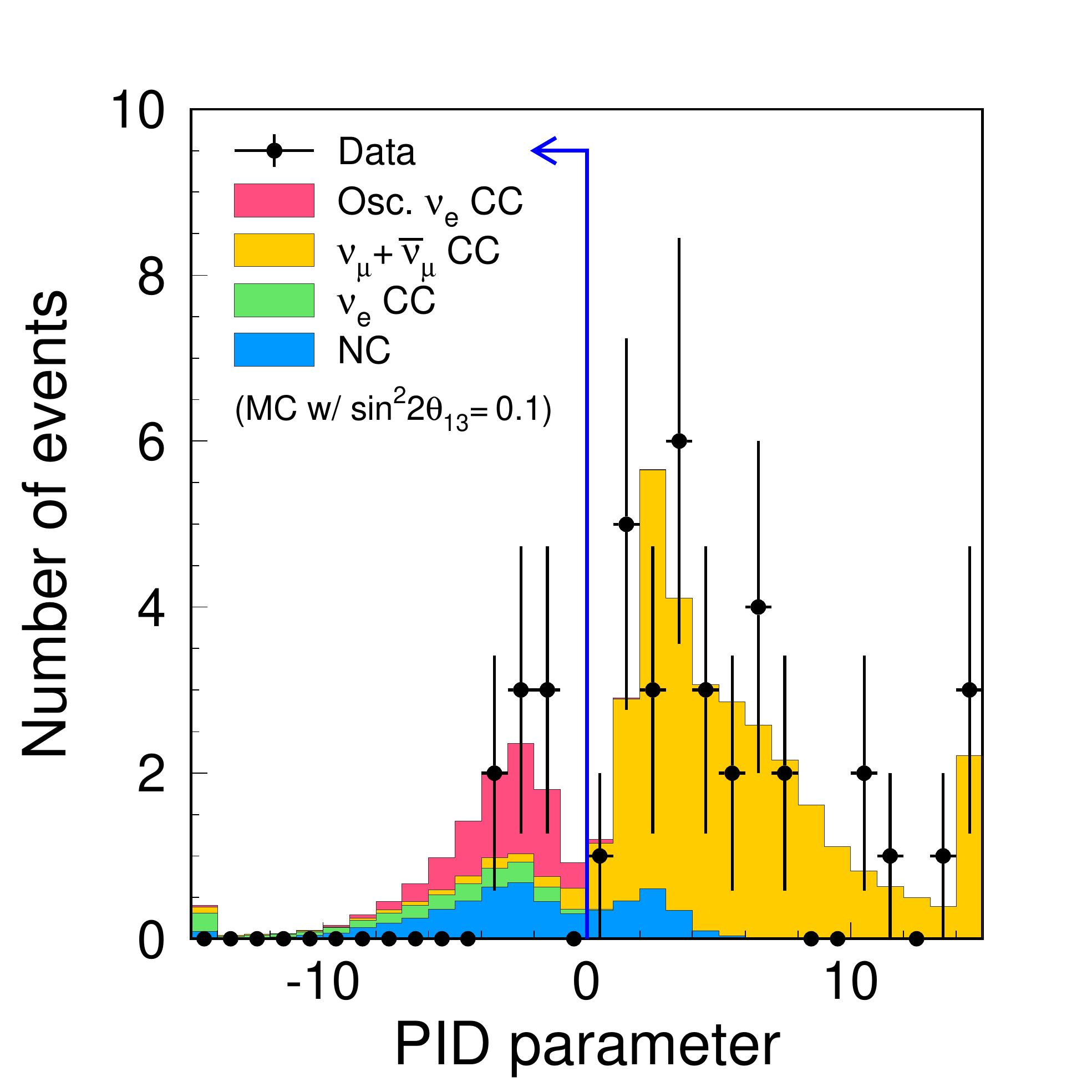}\\
\includegraphics[width=80mm]{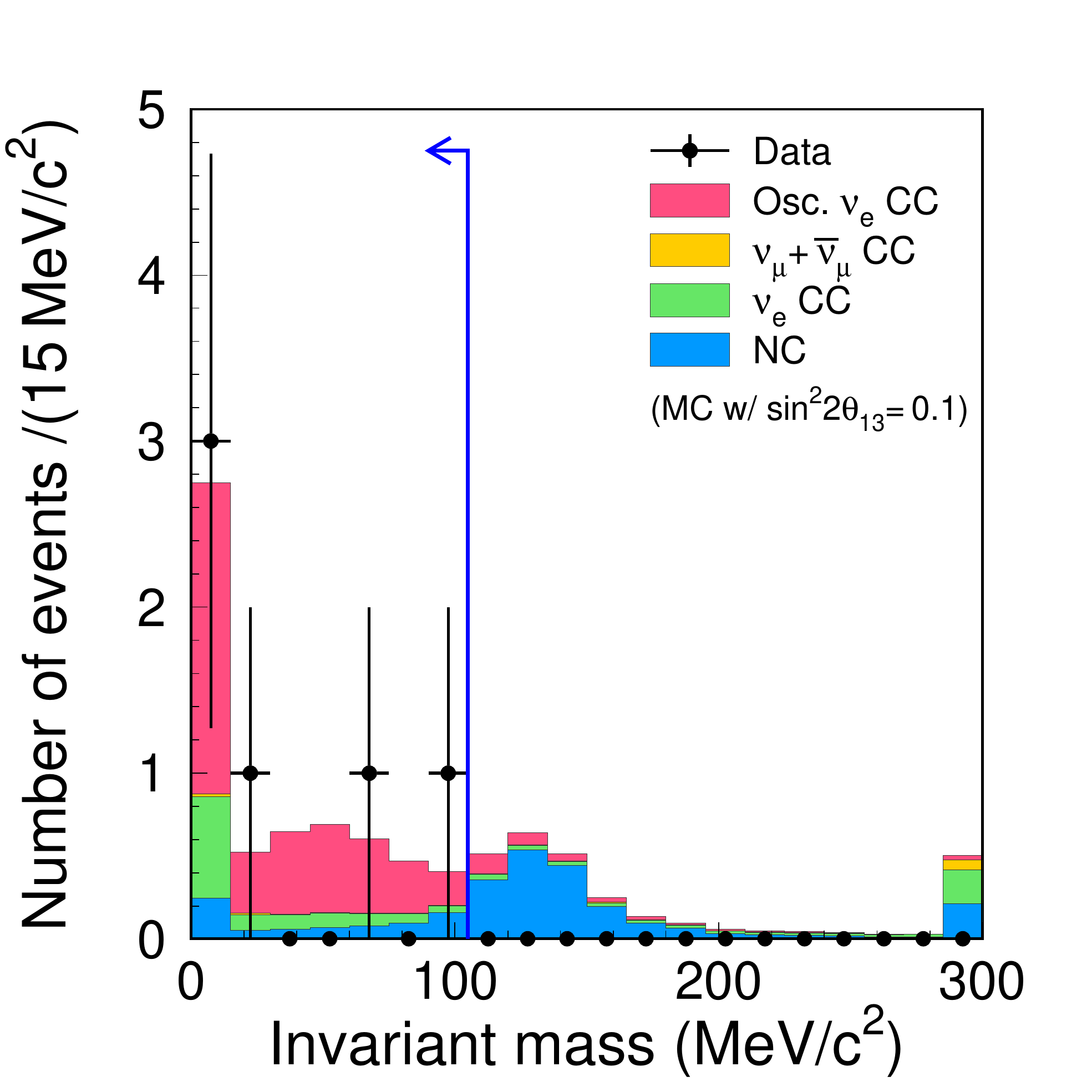}
\caption{Distributions of observables used in the selection of $\nu_e$ candidates in SK. The upper figure shows the particle identification
parameter, which characterizes the sharpness of Cerenkov rings, with sharper rings (muon-like) having larger values. The lower figure
shows the invariant mass for a forced two-ring hypothesis, a powerful statistic to suppress background from neutral current $\pi^0$ production.
The observed numbers of events are compared to expected background and signal if $\sin^2 2\theta_{13}=0.1$.
} \label{fig:sknue}
\end{figure}

Since SK was very well understood and calibrated prior to the start of the T2K program, all $\nu_e$ candidate selection
criteria were set prior to T2K data collection and analysis.
Candidate events were selected from the sample of
fully contained fiducial volume events coincident in time with the pulse of neutrinos from JPARC, having a single ring
pattern consistent with an electron, having visible energy above 100$\,$MeV, having a two-ring invariant mass hypothesis of
less than 105$\,$MeV, and having a reconstructed $\nu_e$ energy less than 1250$\,$MeV.\cite{ref:t2knue}
The distributions of two of the critical quantities in the selection are shown in Fig.~\ref{fig:sknue}.

A total of 6 $\nu_e$ candidate events were selected.
If $\theta_{13}$ is zero, the expected number of selected events is $1.5\pm0.3$.
The probability to observe 6 or more events in this case is 0.007, a p-value that corresponds to a significance of 2.5$\,\sigma$.
This is the strongest indication to date that $\theta_{13}$ differs from zero, and significantly more data is expected from the experiment
in the future.

\subsection{The MINOS experiment}

The MINOS experiment\cite{ref:MINOS} is a long baseline neutrino experiment between Fermilab and the Soudan mine in northern Minnesota
and has been collecting data since 2005.
More information about the experiment is found elsewhere in these proceedings.\cite{ref:sanchez}
The experiment design was not optimized for sensitivity to $\nu_e$ appearance, and has a significant background from
neutral current interactions.
To select a sample enhanced in $\nu_e$ interactions, events are compared to a library of simulated events to find those
with a similar pattern of hits.
A neural network classifier, called LEM, is produced using information from the matching library events.
Because the near and far detector designs are similar,
the near detector can be used to predict the distribution of LEM values for an unoscillated beam in the far detector.

In the MINOS sample, corresponding to $8.2\times10^{20}$ protons on target, a total of 62 events in the far detector
are selected by the $\nu_e$ appearance analysis.\cite{ref:MINOSnue}
If $\theta_{13}$ is zero, the expected number of events is 49.6.
If $\sin^2 2\theta_{13}=0.1$ and $\delta=0$, the expected number of events is 68.2.
The sources of background are neutral current (34.1 events) misidentified charged current (8.8 events) and
intrinsic beam $\nu_e$ interactions (6.2 events).

\begin{figure}
\includegraphics[width=80mm]{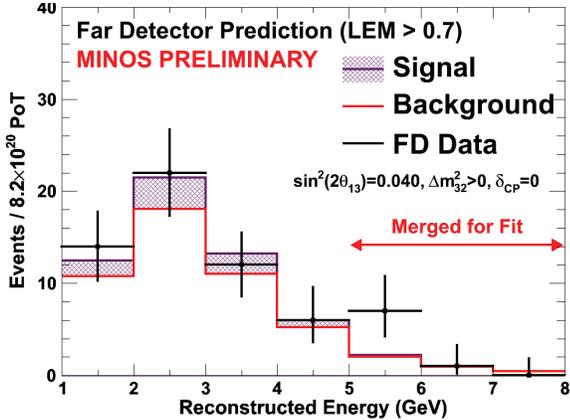}
\caption{Distribution of reconstructed neutrino energies for the selected $\nu_e$ candidate events 
in the MINOS far detector compared to background and
the best fit value of $\sin^2 2\theta_{13}=0.04$ for $\delta_{\rm CP}=0$ and normal hierarchy.
} \label{fig:minosnue}
\end{figure}

The distribution of reconstructed neutrino energies and LEM values are fit in $5\times3$ bins to estimate the value $\theta_{13}$ and
determine a confidence interval.
Figure~\ref{fig:minosnue} shows the reconstructed neutrino energy distribution for the 62 selected events, compared to the
distributions for background only and for the best fit. 
The p-value for the $\theta_{13}=0$ hypothesis is 0.11, and therefore $\theta_{13}=0$ is included in the MINOS 90\% C.L.\ interval.

\subsection{The NOvA experiment}

NOvA\cite{ref:NOVA} is a next generation long baseline experiment between Fermilab and Ash River, Minnesota, using an
off axis angle of 14$\,$mrad to maximize the neutrino flux in the region of 2 GeV, 
appropriate for the baseline distance of 810$\,$km.
The far detector hall has been completed and the construction of the large 15$\,$kton segmented liquid scintillator
detector is expected to be complete in 2014.
The appearance probability, $P(\nu_\mu\rightarrow\nu_e)$, is more sensitive to the CP violation parameter and the neutrino mass ordering with
higher neutrino energies.
As a result, NOvA and T2K will provide complimentary information to the understanding of neutrino oscillations.

\section{SUMMARY}

\begin{figure}[t]
\includegraphics[width=80mm]{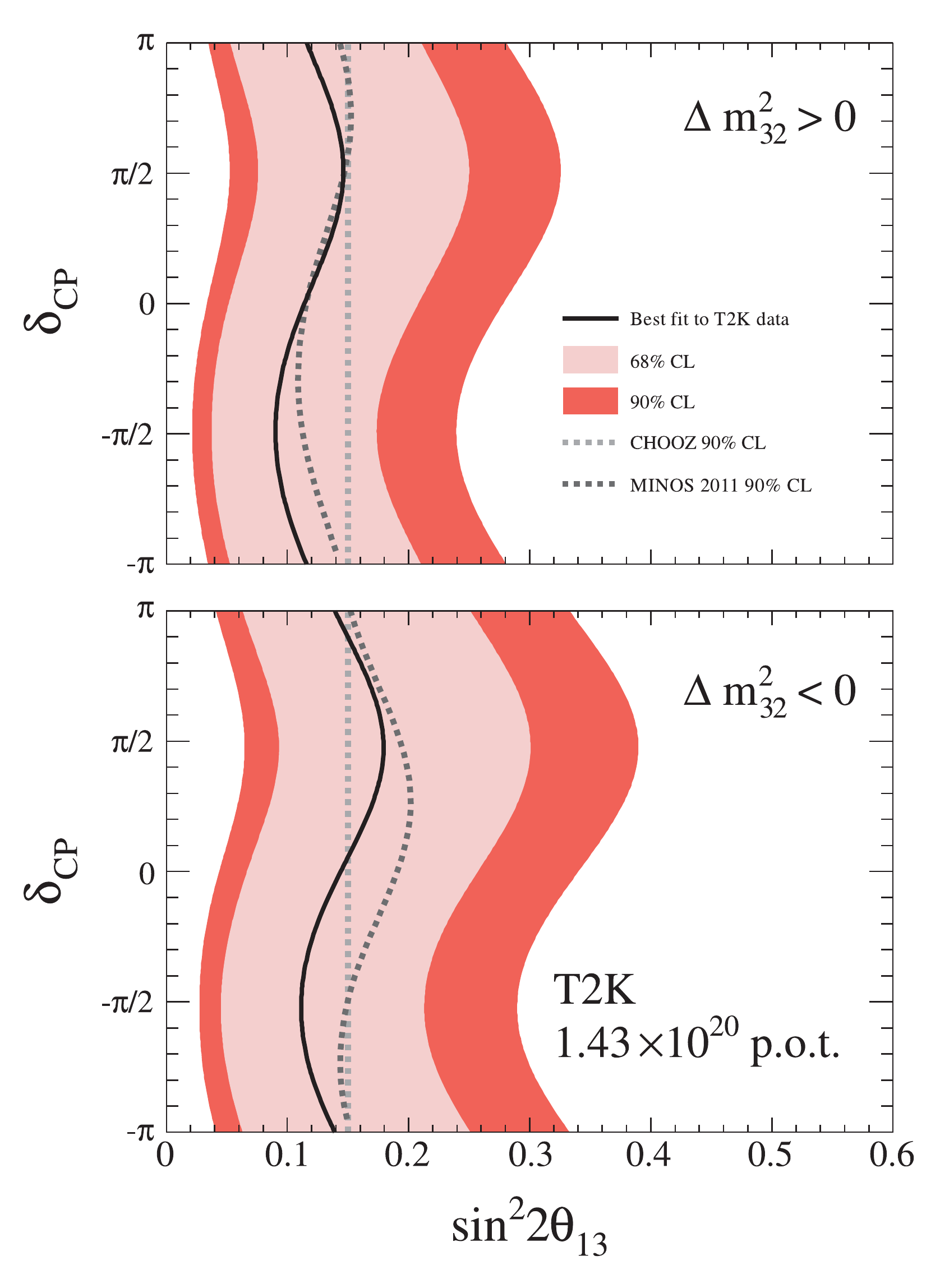}
\caption{Confidence intervals for $\sin^2 2\theta_{13}$ for the normal mass hierarchy (upper)
and inverted hierarchy (lower) as a function of the CP violation parameter $\delta_\textrm{CP}$.
The 90\% CL upper limits from CHOOZ and MINOS are shown by the dashed lines.
The shaded regions show the T2K 68\% and 90\% CL intervals, and the solid line shows the T2K best fit.
} \label{fig:t13}
\end{figure}

There has been significant advances in the world-wide effort to measure $\theta_{13}$ in the past few years.
Precision reactor experiments have come on-line this year and accelerator experiments are giving first indications
that $\theta_{13}$ differs from zero.
A combined analysis of all $\theta_{13}$ measurements has been carried out, and the significance for $\theta_{13}>0$ is now more
than 3$\,\sigma$.\cite{ref:Fogli}
Confidence intervals for $\sin^2 2\theta_{13}$ as a function of the
CP violation parameter are shown in Fig.~\ref{fig:t13} for CHOOZ, T2K and MINOS.

In the coming years it will be interesting to see how the different oscillation measurements compare with each other and
to see if they remain compatible within the PMNS framework.
Combining the present generation of experiments may start to give indications for the CP violation parameter and the neutrino mass
ordering.
Provided $\theta_{13}$ is large enough, future projects with higher beam power and larger detectors
may move ahead to definitively measure the amount of 
CP violation in the lepton sector, important input to understanding the origin of matter/anti-matter asymmetry in the Universe.

\bigskip 

\end{document}